\documentclass[11pt,preprint]{aastex}

\newcommand{\simgt}{\hbox{\rlap{\raise 0.425ex\hbox{$>$}}\lower 0.65ex\hbox{$\sim$}}}
\newcommand{\simlt}{\hbox{\rlap{\raise 0.425ex\hbox{$<$}}\lower 0.65ex\hbox{$\sim$}}}
\newcommand{\vvskip}{\vskip0.05in}

\shorttitle{Angular Distribution of GRBs}
\shortauthors{Williams}

\begin{document}   

\title{Angular Distribution of Gamma-ray Bursts and Weak Lensing}

\author{Liliya L.R. Williams and Natalie Frey\altaffilmark{1}}
\affil{Astronomy Department, University of Minnesota, Minneapolis, MN 55455}
\email{llrw@astro.umn.edu, frey@astro.umn.edu}

\altaffiltext{1}{Present address: 
        Physics Department, University of Central Florida, Orlando, FL 32816}

\begin{abstract}
We investigate whether Gamma-Ray Bursts (GRBs) from the Current BATSE Catalog have 
been affected by weak lensing by the nearby large scale structure. The redshift 
distribution of GRBs is believed to be broad, extending to $z\sim 5$, so most 
events can be assumed to be at large redshifts, and hence subject to weak lensing, 
which would betray itself as projected (anti-)correlations between GRB events
and galaxies or clusters that trace the intervening mass. Given the observed 
distribution of GRBs in fluence $f$, and statistical positional error $e$, we 
predict that most subsets drawn from BATSE Catalog will be anti-correlated with 
the foreground structure due to weak lensing, i.e. will show negative magnification 
bias. We find that GRBs are indeed anti-correlated with the APM galaxies 
($z\sim\,0.2-0.3$) in the sense that galaxy density in circles of radii 
$1^\circ-1.5^\circ$ ($15-20 ~h^{-1}$Mpc at $z\sim 0.3$) centered on 
$e\,\simlt\, 1^\circ$ GRBs is about 10\% lower than expected from a random 
distribution; the significance of GRB-APM anti-correlations reaches 99.7\%. 
Cross-correlation between GRBs and distant rich Abell-Corwin-Olowin clusters 
is also negative. Standard cosmological models with $\Omega_m\sim 0.3$, 
$\Omega_\Lambda\sim 0.7$, and matter distribution on large scales following 
observed APM galaxy distribution with the biasing parameter of around 1 are not 
able to reproduce our GRB-APM anti-correlations. We propose a speculative model 
that does account for these anti-correlations as well as positive correlations 
found previously, between QSOs and APM galaxies. We briefly discuss if the proposed 
scheme is in conflict with observations of cosmic microwave background, galaxy 
surveys, cosmic velocity flows, and weak shear lensing.
\end{abstract}


\newpage
\section{Introduction}\label{intro}

Direct identification of X-ray, optical and radio counterparts of long duration, 
$t>2$ sec, Gamma-Ray Bursts (GRB), and hence their host galaxies has recently 
resolved the GRB distance scale controversy: the observed redshifts span a wide 
range, from $z\sim 0.1$ to $z\sim 5$ \citep{rei01}. This suggests a number 
of uses of GRBs as cosmologically distributed probes. Since they are believed 
to be associated with compact remnants of massive stars, 
it has been suggested they be used to trace star formation rate 
obscured by dust \citep{tot99,djo01}, 
star formation rate at very high redshifts \citep{lam00}, 
as probes of the metal enrichment of the interstellar medium \citep{fio01},
intergalactic medium \citep{fio00}, and galactic and 
intergalactic dust at high redshifts \citep{per00}.

Here, we use GRBs as sources for weak lensing by the large scale structure; 
our ultimate goal is to probe the mass distribution on $\simgt\, 10h^{-1}$ Mpc 
scales, at a typical redshift of 0.3.

Unlike weak shear lensing, which is detected through shape distortion of lensed 
{\it resolved} galaxies \citep{mel99,bar01}, weak lensing of unresolved 
{\it point sources}, like GRBs, would manifest itself through the angular 
(anti-)correlation between sources and lenses, the latter being the intervening 
clumpy mass distribution, which is assumed to be traced by visible galaxies 
\citep{fug90,rod94,bar95}. The sign
of the correlations is determined by the slope of the source number counts
in the appropriate redshift interval. For example, bright QSOs have steep 
number counts implying that positive correlations between these and intervening 
galaxies should be expected. 
In fact, optically selected LBQS QSOs, with $z\ge 1.0$, and radio selected 1Jy QSOs, 
with $0.5\le z \,\simlt\, 2.5$, are independently correlated with faint APM galaxies 
on angular scales of $\sim 1^\circ$ \citep{wil98,nor00}.
The linear extent of the structures is $\simgt\, 15h^{-1}$ Mpc at the redshift of
typical lenses, $z\sim 0.1-0.4$.  Radio selected 1Jy sources are also correlated 
with IRAS galaxies \citep{bar94,bar97}, and Zwicky clusters
\citep{sei95} on angular scales of  $1-2^\circ$.

In these studies,
the redshifts of the sources and probable redshift distribution of the galaxies
do not overlap, insuring that physical associations do not contaminate the lensing 
signal. Qualitative signature of these correlations are those of lensing, however,
it is hard to explain the results quantitatively: 
the observed correlations persist to scales of 
$\simgt\, 1^\circ$ with amplitude of $\omega(1^\circ)\sim 0.02$, whereas
$\omega(1^\circ)\sim 0.002$ is expected if we live in a Universe with 
$\Omega_m\sim 0.3$, and matter power spectrum not too different 
from the observed galaxy power spectrum on large scales 
\citep{dol97,san97,bar97}.
If lensing induced correlations were indeed as small as predicted, they could 
not have been detected on $\sim 1^\circ$ scales given the available number 
of sources \citep{bar93,bar97}.

These results by themselves are puzzling. However, they become 
problematic in view of the recent observations of cosmic {\it weak shear} 
lensing, which agree well with the currently accepted cosmological model,
matter fluctuation spectrum, and biasing parameter close to unity.
There are now several independent determinations of
weak shear lensing in fields of up to $30^\prime$ diameter 
\citep{bac00,hoe02,kai00,mao01,van00,wit00}, 
and the agreement between them is impressive.  Furthermore, when cast in
terms of the best estimate for $\sigma_8 {\Omega_m}^{\sim 0.5}$, weak lensing
results agree remarkably well with cluster normalization constraints 
\citep{mel01}.

GRBs provide us with a different set of cosmologically distributed sources,
which should also be affected by weak lensing, just as QSOs are. However, 
as we explain below, unlike QSOs, which are predicted and shown to exhibit 
correlations with galaxies, GRBs, due to their distribution in fluence and 
angular positional error, are predicted to be anti-correlated with the foreground 
lensing mass distribution. Thus, GRBs should provide a new and interesting 
test of weak lensing on large scales.  

There are two potential difficulties in using GRBs in a weak lensing study.

First, GRB positions are not well localized, with errors ranging from a 
fraction of a degree to as high as $\sim 30^\circ$  for some events with 
low fluences. Obviously, correlation scales that can be reliably probed 
are limited by the GRB position errors on the sky. To circumvent this problem
we use GRB subsets with upper limits on error. 

Second, since individual GRB redshifts are largely unknown,
some of the GRBs at low redshifts will be physically associated 
with the galaxies which we use to trace the lensing mass, and will `contaminate' 
the lensing induced signal. 
Currently, only about 26 GRBs, i.e. a very small fraction of all events have 
confirmed redshifts\footnote{{\url http://www.aip.de/\~\,jcg/grbgen.html}, 
site maintained by J. Greiner, Astrophysical Institute Potsdam.}. 
Of these, about 3 are at $z< 0.4$.
Guided by the known redshifts, several workers have proposed empirical
relations connecting observable properties of GRBs, such as functions constructed
out of time information, to their luminosities \citep{ste99,nrr00,rii01}. 
Redshift distribution implied by these indicate that the true redshift distribution 
is very broad, extending to $z\sim 5-10$.
The fraction of low redshift events is expected to be small;
for example, using the model of \citet{rei01} and their Fig. 3 
histogram, the fraction of events below $z_s=0.2\,(0.4)$ is roughly estimated 
to be 6\% (13\%). As tracers of dark matter on large scales we use
APM galaxies and Abell-Corwin-Olowin (ACO) clusters, which peak at $z\sim 0.2$, 
and extend to a maximum $z=0.4$. So the fraction of GRBs arising in these 
structures should be small, $\simlt 10\%$. Furthermore, and more importantly, we 
predict and find anti-correlations between GRBs and intervening 
galaxies, and so physical associations, if any, would have diminished the amplitude 
of lensing induced anti-correlations. Thus our observed signal (Section~\ref{APM})
is a lower limit. As we show in Section~\ref{can_we?}, the observed amplitude of 
lensing induced anti-correlations is hard to explain within standard
cosmological models, so at this point there is no motivation to carefully 
subtract the effect of physical associations from the observed anti-correlation
signal; we neglect the effect of physical associations in our analysis.

The plan of the paper is as follows.  
After describing the data in Section~\ref{data} we use the fluence and error 
information of Current BATSE Catalog GRBs to predict the amplitude and sign of 
GRB-galaxy correlations, i.e. compute magnification bias (Section~\ref{underden}). 
In Sections~\ref{APM} and ~\ref{ACO} we estimate the strength of observed 
correlations between GRBs and APM galaxies and ACO clusters, respectively.
Anti-correlations are found in both cases, as predicted, but the amplitude 
of the effect is higher than expected. In Sections~\ref{we_can} we discuss the 
various possibilities for reconciling observations with theory, and propose a 
speculative scenario. We summarize and discuss our findings in 
Section~\ref{sum+disc}.

\section{Data selection}\label{data}

\subsection{Gamma-Ray Bursts}\label{GRBs}

We use the Current BATSE GRB Catalog, ending with trigger number 8121, which
occurred on May 26, 2000. From these events we select those that were not 
overwritten by a later more intense trigger,
and that have non-zero fluences in the 50-100 keV and 100-300 keV energy 
channels. GRB fluxes and fluences are recorded in 4 channels, which cover energy 
ranges 20-50 keV, 50-100 keV, 100-300 keV, and $> 300$ keV, respectively.
We use the fluences in the middle two channels %(Section~\ref{underden})
because the corresponding energy range has the peak flux, and coincides with 
the energy range of the nominal BATSE on-board burst trigger
({\url http://www.batse.msfc.nasa.gov/batse/grb/catalog/4b/4br\_flux.html}).
These cuts leave us with 2038 events.

BATSE events are not well localized in the sky;
positional errors come in two flavors, statistical and systematic.
Statistical errors, $e$, are recorded in the BATSE Catalog, and range from a
fraction of a degree to $30^\circ$ degrees. 
(All quoted errors are $\sqrt{e_x^2+e_y^2}$.)
The peak of the statistical error 
distribution is at about $3^\circ$. Systematic errors, $e_{sys}$, 
were analyzed for the revised 4B Catalog \citep{pac99}, and 
were found to have a modified Gaussian distribution such that 78\% of
the events have errors of $\leq 1.85^\circ$. For our purposes an 
approximation used for the 3B Catalog will suffice: we assume the systematic
error distribution to be a Gaussian such that 68\% of the events have 
systematic errors of $\leq 1.6^\circ$. 
Assuming that statistical and systematic errors are independent, the total 
error is $e_{tot}=\sqrt{e_{sys}^2+e^2}$. The statistical errors and fluences
of the 2038 bursts are plotted in Fig.~\ref{ncounts}. Solid squares represent
10 BATSE GRBs with known fluences and redshifts. 

BATSE Catalogs have a declination dependent completeness level on the sky, 
with a 74\% difference between the most ($\delta=75^\circ$) and least 
($\delta=-10^\circ$) completely covered declinations \citep{pac99}.
This incompleteness will not matter for our analysis in Section~\ref{APM}, 
because we compare galaxy density around each GRB with the expected average 
density on the same APM plate. This incompleteness
will be important in the cross-correlation with ACO galaxy clusters, 
Section~\ref{ACO}, and will be accounted for.

\subsection{APM Galaxy Catalog}\label{gal+clus}

APM Galaxy Catalog \citep{irw94}
is digitized POSSI plate material, with magnitude limits 
20.0 and 21.5 on `red' and `blue' POSSI plates in the Northern Hemisphere, and
21.0 and 22.5 on `red' and `blue' UKST plates in the Southern Hemisphere,
with internal accuracy of $\sim 0.1$ mag for all but the faintest objects,
and external, i.e. plate-to-plate accuracy of $\sim 0.3$ mag.
These are tolerable errors for our purposes: internal accuracy is sufficiently
good to trace the light fluctuations on each plate, and external accuracy is
less important to us because we confine analysis for each GRB
to a single APM plate. Similar to previous work \citep{wil98} we
select galaxies as objects that are classified as extended on the `red' plates 
by the APM. The magnitude range is set to probe the most distant regions
reached by the APM: 
on POSSI, $18.5 \leq m_R \leq 20.0$, and 
on UKST,  $19.5 \leq m_R \leq 21.0$.

In the Northern Hemisphere APM plates do not cover a band between $-20^\circ$ 
and $+20^\circ$ Galactic Latitude, and in the Southern Hemisphere
the coverage is even less complete. Furthermore, even though each plate
is $5.8^\circ\times5.8^\circ$, because of plate vignetting
we do not use GRBs located beyond $2.5^\circ$ from the plate center.
Thus, we only use a subset of 732 GRBs in the APM analysis in Section~\ref{APM}.

\section{Predicting the effect of Weak Lensing on the GRBs}\label{underden} 

The intervening mass distribution between us and GRB events is an uneven
lens that stretches some areas of the background sky and shrinks others. 
The redshift distribution of the observed APM galaxies is such that they trace
dark matter from $z_{min}\sim 0.1$ to a maximum of $z_{max}\sim 0.4$.
(see Fig. 1 of Williams \& Irwin 1998). Let the total average lensing optical depth
of this slab of matter be $\kappa_0$:
\begin{equation}
\kappa_0=\rho_{crit} \Omega_0
\int_{z_{min}}^{z_{max}} {{(c dt/dz) (1+z)^3}\over{\Sigma_{crit}(z,z_s)}}dz.
\label{tau}
\end{equation}
Here, $cdt$ is the thickness of the lensing slice at redshift $z$,
$\rho_{crit} \Omega_0 (1+z)^3$ is its mass density, and
$\Sigma_{crit}(z,z_s)$ is the critical lensing surface mass density
at $z$ for a source at $z_s$. Within this redshift range, the 
fluctuations in the projected surface mass density are $\delta\sigma/\sigma$,
and are assumed to be small on large scales.
For a specified cosmology and source and lens redshifts, these fluctuations are 
translated into (de-)magnifications on the sky, with respect to the 
`smooth Universe' case: 
$M=(1-\kappa)^{-2} \approx 1+2\kappa$, where 
$\kappa=\kappa_0~(\delta\sigma/\sigma)$.

For sources with exactly known positions, like QSOs, this magnification field 
$M$, combined with the QSO number counts for the relevant range of redshifts, 
can be translated into an observed distribution of number density on the sky 
of sources down to some specified flux limit. 
If $\alpha=d\,{\rm log}\,n_{QSO}(<m)/dm$ is the slope of the number counts near
the survey flux limit, then the number density of QSOs will be a factor of 
$q=M^{2.5\alpha-1}$ different from the `smooth Universe' ($M=1$) case; 
$q$ is called source over- or underdensity, depending on whether it is 
greater or less than 1. The fact that it is different from 1 is called 
the magnification bias. Bright QSOs, whose number counts are steep, would appear 
to be $q$ times more abundant in the directions of mass concentrations,
i.e. they would appear to be correlated on the sky with the nearby structure.
Faint QSOs, on the other hand, have a shallow number counts slope, 
and so will be anti-correlated with the foreground lenses. 

\begin{figure}
\plotone{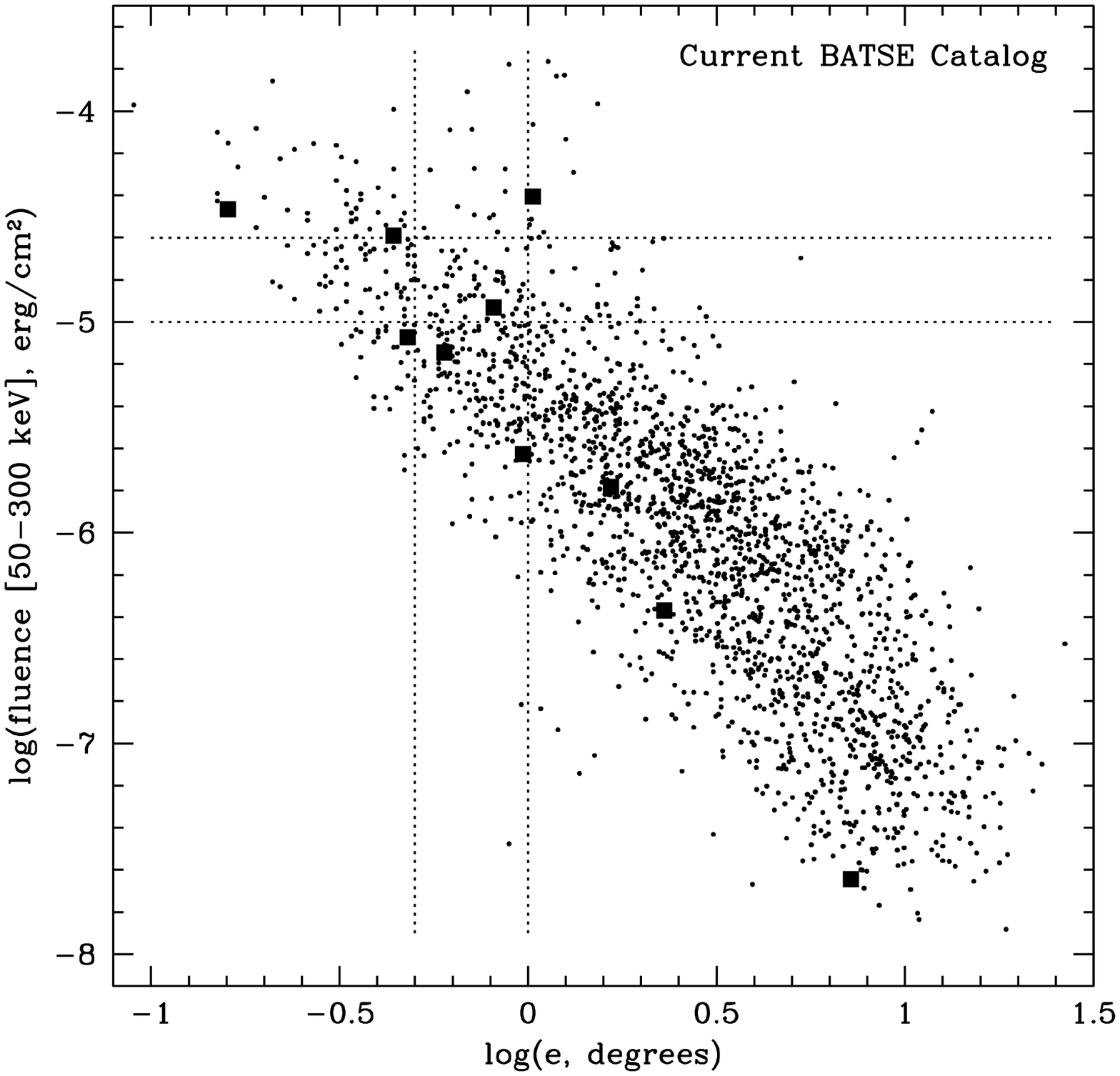}
\caption{
Distribution in fluence and statistical position error of 2038 GRBs from 
the Current BATSE Catalog. The fluence is the sum of Channels 2 and 3, 
corresponding to the $50-300$ KeV range. Filled squares represent 10 BATSE
GRBs with known redshifts and fluences. The four dotted lines delineate the 
boundaries of the four GRB subsets considered in Section~\ref{underden}.
\label{ncounts}}
\end{figure}

With GRBs, the calculation of $q$ is somewhat different, because in addition
to limiting these in fluence $f$, one also needs to place an upper limit 
on GRB positional error $e$, so that the GRBs have a reasonable chance of
being within the specified area of correlations. 
Even though positional errors are not 
directly affected by lensing, the observed GRB errors are
well correlated with their fluences, therefore, as a GRB's fluence is 
changed through magnification, so is its error. Thus,
in order to correctly predict $q$ one needs to consider 
GRB distribution in $e$ and $f$, Fig.~\ref{ncounts}. 

Let $p(e|f)$ be the normalized probability distribution of GRB sources in 
statistical positional error $e$ at a given fluence $f$. As $f$ we use the sum 
of fluences in Channels 2 and 3.
Probability distribution $p(e|f)$ is estimated from the BATSE data itself,
by binning the distribution presented in Fig.~\ref{ncounts} by fluence and error, 
and then normalizing distributions in $e$ for each fluence separately.
If the sources are limited in error by $e_1$ and
$e_2$ from below and above, and in fluence by $f_1$ from below, then the 
number of such sources seen behind a smooth patch of lens with magnification 
$M$ is given by,
\begin{equation}
n_{GRB}(e_1,e_2,f_1,M)
      ={1\over M}\int_{f_1}^\infty \Big[ n_{GRB,0}({f^\prime/M})
      ~\int_{e_1}^{e_2} p(e|f^\prime)~de\Big]~df^\prime,
\label{Nlensed}
\end{equation}
where $n_{GRB,0}(f)$ is the distribution of sources in fluence one would see in a 
Universe with completely smooth mass distribution. This distribution is not 
observable; however, given the small typical magnifications it is reasonable 
to assume that the observed sources give a fair representation of $n_{GRB,0}(f)$.

The ratio of the number of sources observed and the number
that would be seen if the mass were smoothly distributed everywhere,
is the over- or underdensity,
\begin{equation}
q(e_1,e_2,f_1,M)={{n_{GRB}(e_1,e_2,f_1,M)}\over{n_{GRB}(e_1,e_2,f_1,M=1)}}
\label{q_eqn}
\end{equation} 

\begin{figure}
\plotone{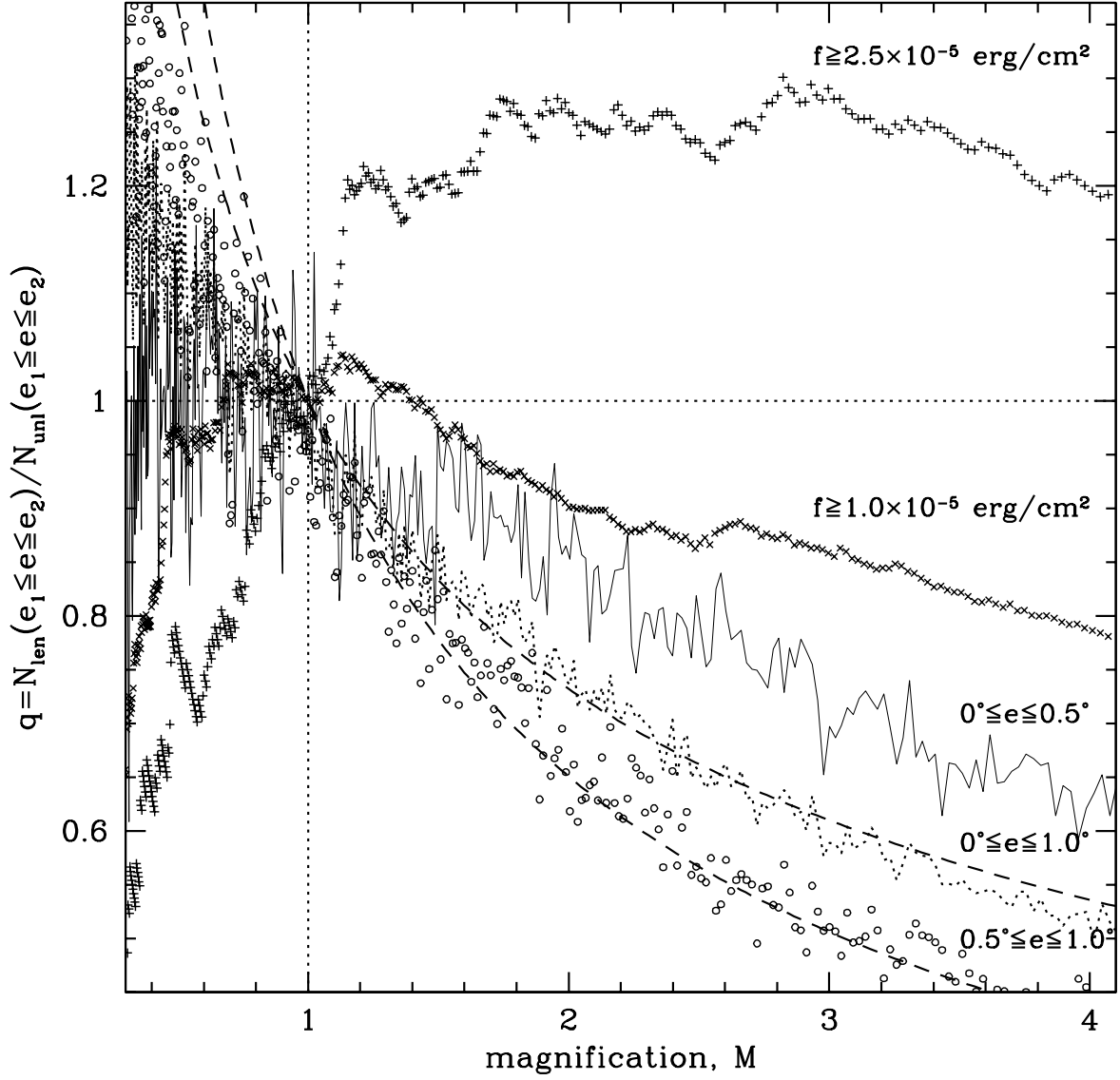}
\caption{Computing magnification bias for BATSE GRBs from Fig.~\ref{ncounts}: 
$q$ is the predicted overdensity of GRBs in the direction of a 
patch of intervening matter with a constant magnification $M$.
See Section~\ref{underden} for details.
\label{overden}}
\end{figure}

Figure~\ref{overden} shows the result, for a few combinations of $e_1$,
$e_2$, and $f_1$. The upper two lines (cross and plus symbols) are for GRB
subsets limited in fluence only, with no restrictions imposed on error.
The lower three lines are for GRB subsets limited in error, with no limits
placed on fluence.  We used all 2038 GRBs in the Current BATSE Catalog to 
construct the $(M,q)$ relation in Fig.~\ref{overden}.
Had we used, say the 732 GRBs that are found on APM plates the shape of
the curves above would have been the same, within the noise. Similarly, 
the plot is not very sensitive to the particular choice of fluence channel.
Note that all lines go through $(M,q)=(1,1)$, as they should. 
Some of the $q(M)$ lines vary non-monotonically with magnification, 
and most lines vary erratically. This is due to the noise associated with 
the finite number of GRB points in Fig.~\ref{ncounts}. Shot noise is especially 
pronounced when GRBs with small errors 
and/or large fluences are considered, and when $M<1$: in that 
case the $n_{GRB,0}$ term in eq.~\ref{Nlensed} refers to a very small number 
of GRBs at highest fluences, and so shot noise fluctuations can make the 
resulting $q$ vary a lot. 

Most of the subsets that can be constructed from the total GRB catalog using 
different fluence/error cuts are predicted to be anti-correlated with lenses. 
In general, overdensities would be expected only if there is a large `reservoir' of 
sources just below the detection limit. The fact that the predicted GRB $q$'s tend 
to be $<1$ can be seen directly from Fig.~\ref{ncounts}: for almost any $e$, $f$ cut
there is no large `reservoir' of sources just below the $e$, $f$ limits ready
to be magnified into the observed subset.
The only subsets of GRBs which are predicted to be correlated with lenses
are those with large fluences, $f_1\simgt 2.5\times 10^{-5}$ erg/cm$^2$.
However, these subsets contain very small numbers of GRBs, 
so the uncertainties in the $(M,q)$ relation for these are probably large. 

Dashed lines in Fig.~\ref{overden} are fits to the 
$(e_1,\,e_2,\,f_1)$=$(0.0^\circ, 1.0^\circ, 0.0)$ and $(0.5^\circ, 1.0^\circ, 0.0)$
GRB subsets, which we will be considering in some detail in the next Section.  
We wanted fits of the form $q=M^{\beta}$, to mimic the QSO relation, 
$q=M^{2.5\alpha-1}$. In the present case, $\beta=-0.62$ and $-0.45$ provide 
adequate fits for the two subsets respectively; the corresponding $\alpha$'s 
are $0.152$, and $0.22$. The fits are only rough,
but are completely adequate for our purposes.

Having made predictions as to the amplitude and sign of the GRB correlations
with the foreground matter, we can now proceed to do the corresponding 
`observations'.

\section{GRB--APM galaxy correlations}\label{APM}

In principle, we want to determine the number density of GRBs
as a function of projected mass excess, $\delta\sigma/\sigma$. However,
since GRBs are rare and galaxies are plentiful, we instead estimate galaxy
density in circles around GRBs. Our analysis later, Section~\ref{can_we?}
will take this difference into account.

\begin{figure}
\epsscale{0.9}
\vskip-2.25in
\plotone{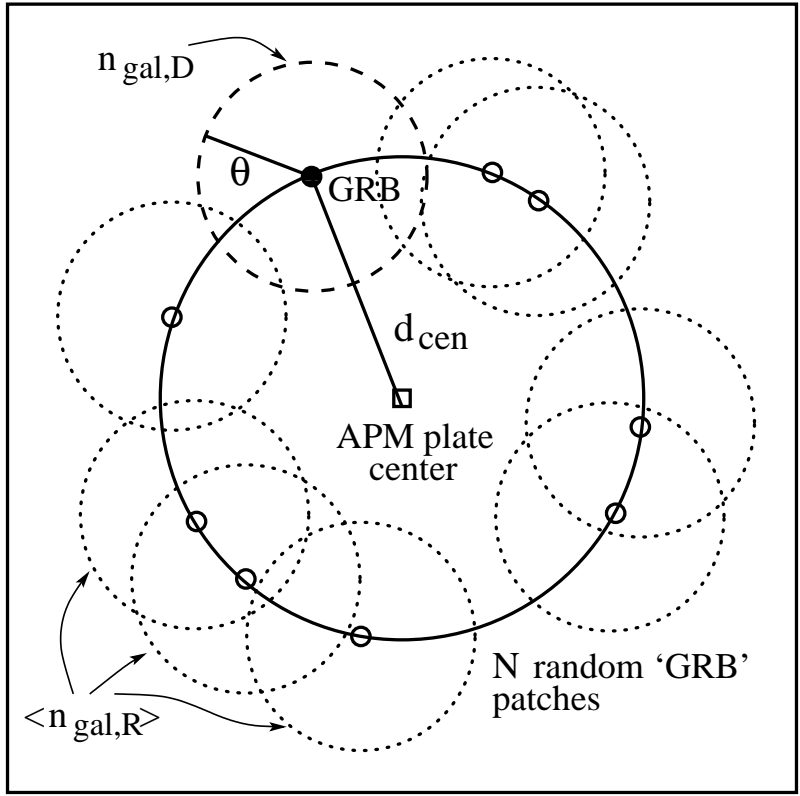}
\vskip-1.5in
\caption{
Schematic of an APM plate that demonstrates how we choose control patches.
$d_{cen}$ is the distance of GRB from plate center, and $\theta$ is the radius 
of the patch inside of which we count galaxies. $n_{gal,D}$ is the number of 
galaxies inside the GRB patch (dashed circle), while $\langle n_{gal,R}\rangle$ 
is the average over random patches (dotted circles). The random patches
are at the same distance away from the plate center as the GRB itself.
This particular placement of random control patches is designed to take care
of vignetting and radial object density variations on APM plates. GRBs that are 
not within $d_{cen}=2.5^\circ$ of any APM plate are not used. 
When $\theta$-patches run off the edge of the plate the areas that fall outside 
the plate are not used. So a $\theta=1.5^\circ$ circle around a GRB close to 
plate center will have a larger area than a $\theta=1.5^\circ$ circle around 
a GRB on the edge of a plate. This difference is accounted for by placing
the random GRBs at the same plate-centric distance as the original GRB.
\label{schematic}}
\epsscale{1.0}
\end{figure}

Let the number of galaxies within $\theta$ degrees of a GRB, and
in the magnitude ranges specified in Section~\ref{gal+clus}, be $n_{gal,D}$.
The latter should be normalized by average
expected galaxy count within similarly sized random $\theta$-patches on the sky.
In our analysis, because the APM plate-to-plate magnitude calibration can be
rather uncertain, about 0.3 mag, we select control areas for normalization
from the same plate as the corresponding GRB. Furthermore, because the
object density on plates varies as a function of distance from center, the
control patches are restricted to lie at the same distance from plate
center as the GRB, $d_{cen}$. Fig.~\ref{schematic} illustrates our selection 
of control patches. For each GRB, $N$ such random positions are created. We
scale $N$ linearly with $d_{cen}$, such that at $d_{cen}=1^\circ$, $N=100$. (We 
tried using same $N$ for all GRBs; the results below did not change substantially.)
The number of 
galaxies in $\theta$ circles around these are also recorded, and the average
is calculated for every GRB, $\langle n_{gal,R}\rangle$. 
The galaxy excess, $[n_{gal,D}/\langle n_{gal,R}\rangle]-1$, is then equal to 
$b~(\delta\sigma/\sigma)$, where $b$ is the biasing parameter of APM galaxies 
with respect to the underlying mass.

\subsection{Two GRB subsets most likely to show weak lensing effects}\label{one}

Using Fig.~\ref{overden} we select GRB subsets that are most likely to show 
weak lensing induced signature. We choose subsets with 
$(e_1, e_2)=(0.5^\circ, 1.0^\circ)$ and $(0^\circ, 1.0^\circ)$.
How do we decide on the size of the $\theta$-patch around each GRB? 
Obviously, each GRB should
have a good chance of actually being located within these circles. 
The limiting factor here is the systematic positional error: even if
a GRB has zero statistical error there is only a 25\% chance that it will
be actually within a $0.5^\circ$ radius drawn around it. 
On the other hand, the size of the APM plate limits the size of $\theta$ from
above; we settle on $\theta=1.5^\circ$. 

Given that the GRB and the $N$ control $\theta$-patches are equidistant from 
plate center, to avoid significant overlap between these 
we restrict our GRB subset further by selecting only those at $d_{cen}\geq\theta$. 
Thus GRBs are limited to lie at $1.5^\circ\leq d_{cen}\leq 2.5^\circ$. 
The $(e_1, e_2)=(0.5^\circ, 1.0^\circ)$ and $(0^\circ, 1.0^\circ)$ subsets
contain 46 GRBs and 74 GRBs respectively, and we will refer to them by quoting 
these numbers. 

We now ask if the galaxy density around GRBs in the two subsets is 
different from what one would expect if random points were used in 
place of GRBs. Instead of generating a set of random points for this purpose, we
use the whole set of GRBs found at $1.5^\circ\leq d_{cen}\leq 2.5^\circ$, 
regardless of positional error. There are 448 of these. Since most of these
have rather large errors, their real positions are quite far from BATSE
recorded positions, so in effect we have a set of randomly selected points.

In Fig.~\ref{subsample_hist} the heavy solid histogram is the distribution
of $n_{gal,D}/\langle n_{gal,R}\rangle$, or, equivalently, 
$b~(\delta\sigma/\sigma)+1$ for the 448 `random' points. 
The dashed and dotted histograms are the 46 and 74 GRB subsets respectively. 
The averages of the three distributions are 0.998, 0.883, and 0.915,
so GRBs in the two subsets are found in the directions of 
foreground regions that are, on average, 12\% and 8\% underdense in galaxies. 
In other words, we detect anti-correlations between GRBs and intervening galaxies,
which is what was predicted in Section~\ref{underden} and Fig.~\ref{overden}. 
We leave quantitative comparison with predictions until Section~\ref{can_we?}.

Let us now evaluate the statistical significance of this result.
For each GRB we calculate $N_{>}/N$, 
where $N_{>}$ is the number of random $\theta$-patches, out of total $N$, 
that have less galaxies in them than the $\theta-$patch around the real GRB. 
In other words,
$N_{>}/N$ is the rank of the real GRB patch among its `random peers';
$N_{>}/N=0.5$ if GRBs are randomly distributed with respect to
the foreground galaxies, but if GRBs have an excess of galaxies in their 
foregrounds then $N_{>}/N>0.5$. If GRBs occupy
random positions with respect to the foreground galaxies, then the
{\it distribution} of $N_{>}/N$ values is known---the cumulative 
distribution should be linear, and in fact for the whole set of 448 GRBs it is,
see solid line in Fig.~\ref{smkstest2_histo}. The dashed and dotted histograms
are for the 46 and 74 GRB subsets, respectively; their $N_{>}/N$ values are
0.381 and 0.423, implying that GRBs have a deficit of galaxies in their
foregrounds. Using the Kolmogorov-Smirnof (KS) test these two distributions 
differ from the
whole set of 448 GRBs at the 99.68\% and 97.62\% significance level.
Note that the 46 subset [$(e_1, e_2)=(0.5^\circ, 1.0^\circ)$] was predicted 
to be more strongly anti-correlated with galaxies than the 74 subset
[$(e_1, e_2)=(0.5^\circ, 1.0^\circ)$], which is what is observed, in spite
of the 46 subset having less GRBs than the 74 subset.

Since the 46 and 74 GRBs in the two subsets are a part of the 448 `random' 
points, these results could be a conservative estimate of statistical 
significance. However, if we use truly random points, the statistical 
significance of the results is not much higher.

\subsection{Many GRB subsets}\label{many}

A further test of the statistical significance of the results for the
two subsets in the last Section is provided by carrying out the same 
analysis on many GRB subsets, selected using
a range of error and fluence criteria.
Subsets are defined by $\theta$, the size of the patch around GRBs, 
statistical error range, $e_1\rightarrow e_2$, and the range of GRB's $d_{cen}$. 
We use four different $\Delta e$ ranges:
$0.5^\circ$, $1.0^\circ$, $1.5^\circ$, and $2.0^\circ$, and include GRBs with
errors up to $10^\circ$. 
For each of these we use five different $d_{cen}$ ranges:
$0.0^\circ-2.5^\circ$, $0.5^\circ-2.5^\circ$, ...., $2.0^\circ-2.5^\circ$. 
We try three values of $\theta$: $0.5^\circ$, $1.0^\circ$, and $1.5^\circ$.
The total number of GRB subsets is 615; the
number of GRBs in these subsets varies from 1 to 269. 

Of the 615 total cases considered, Table 1 lists those that deviate from
the parent GRB set at more than 97\% confidence level, as judged by the KS test.
The two cases considered in Section~\ref{one} are marked with a star. Most of 
the other cases in Table 1 are related to these two subsets. Even though the
subsets are not independent, one should still expect some small fraction of the 
615 subsets to be significant at $>97\%$, so it is not surprising that we have these.
It is interesting, however, that the subsets that do show significant deviations 
form the parent GRB population
are the ones that we argued would be most likely to be affected by lensing.
Furthermore, all cases reported in Table 1 are {\it anti}-correlations, consistent
with weak lensing predictions of Section~\ref{underden}.

\subsection{Can we account for the GRB-galaxy anti-correlations?}\label{can_we?}

\begin{figure}
\plotone{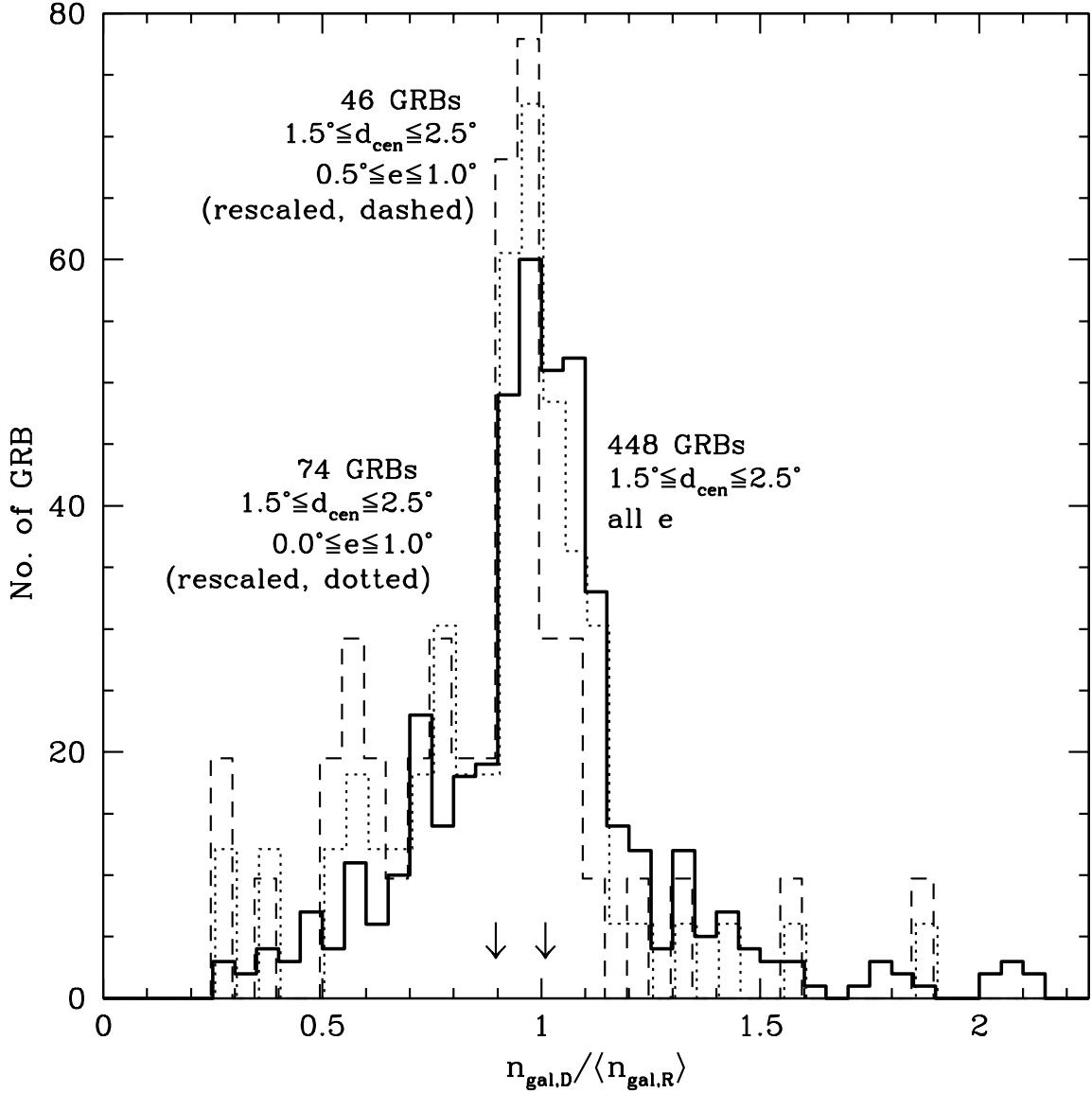}
\caption{Histograms of normalized projected galaxy number density, 
$n_{gal,D}/\langle n_{gal,R}\rangle$ in circles of radius $\theta=1.5^\circ$
in the directions of GRBs. The dashed line is for 46 GRBs with 
$0.5^\circ\le\,e\le\,1.0^\circ$, while the dotted line is for 74 GRBs with
$0^\circ\le\,e\le\,1.0^\circ$. These two histograms have been scaled to
match the area under the curve of the solid line histogram, which
represents 448 GRBs used as the control set. All GRBs are limited to 
$1.5^\circ\le\,d_{cen}\le\,2.5^\circ$ from APM plate center.
The arrows indicate the averages for the three distributions.
GRBs in the 46 and 74 subsets have a deficit of APM galaxies in front of them.
\label{subsample_hist}}
\end{figure}

\begin{figure}
\plotone{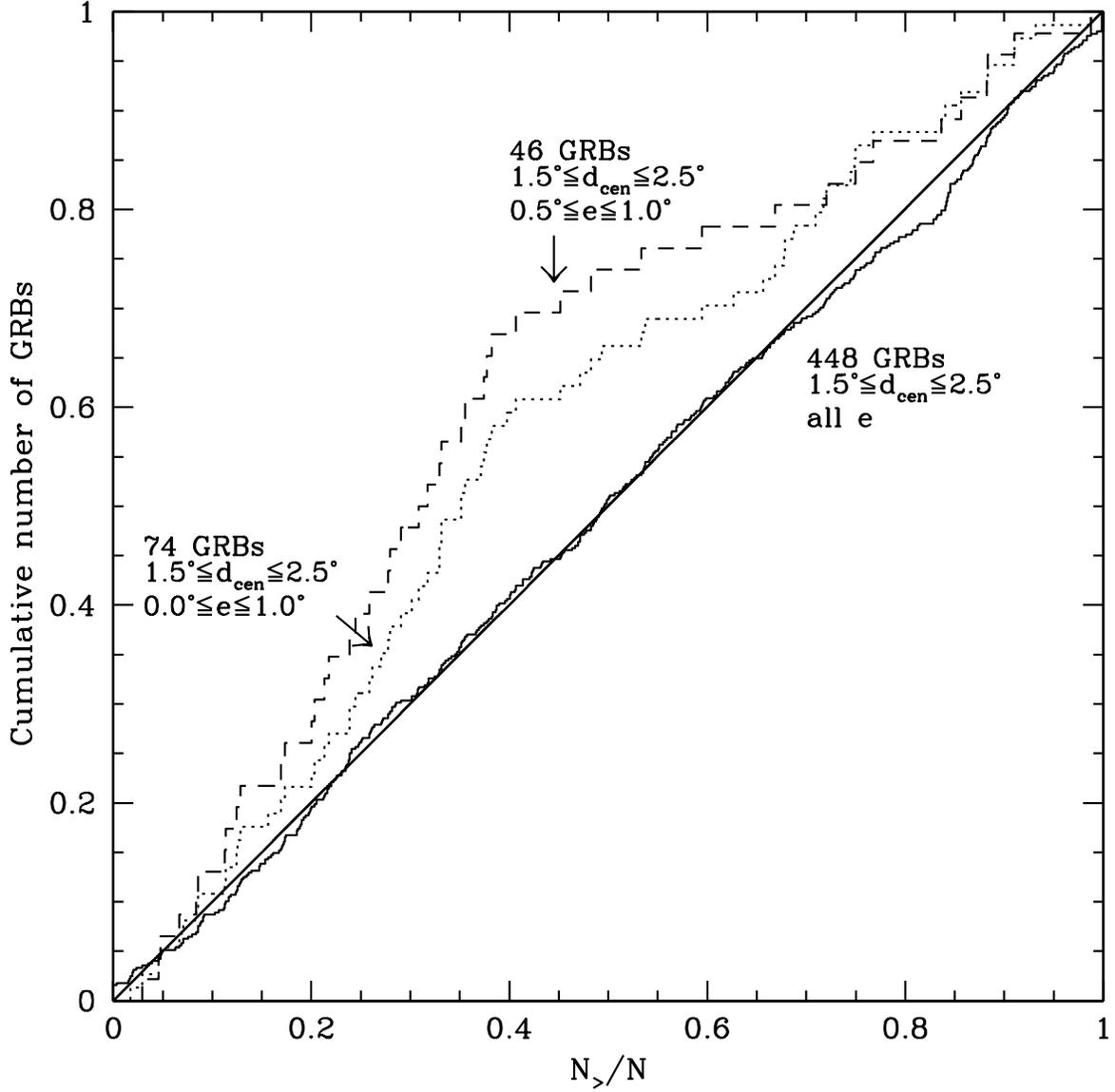}
\caption{The cumulative distributions of $N_{>}/N$ for the control GRB
subset of 448, and the 46 and 74 subsets. $N_{>}/N$ is the fraction of random 
$\theta-$patches that have less galaxies than the GRB-centered patch. In the case 
of no correlations the distribution should follow a diagonal line (dotted).
According to the KS 2-sample test the 46 and 74 subsets could not have been drawn 
from the 448 set at 99.7\% and 97.6\% confidence levels, respectively. The whole set 
of 448 GRBs is not significantly different a random distribution of $N_{>}/N$ values.
\label{smkstest2_histo}}
\end{figure}

Here we attempt to account for the amplitude of anti-correlations of two GRB
subsets considered in Section~\ref{one}.
As before, we assume that the whole set of 448 GRBs, found at distances
$1.5^\circ\leq d_{cen}\leq 2.5^\circ$ from plate centers is a collection of 
random points on the sky, since most of them have very large position errors. 
Hence, the solid line histogram of Fig.~\ref{subsample_hist} is a fair
representation of the distribution of average projected galaxy densities in
$\theta=1.5^\circ$ circles. This counts-in-cells distribution does not 
include fluctuation power on scales larger than the size of an APM plate.
Because we are dealing with weak lensing regime, we can separate the effects of 
density fluctuations on different projected scales; our `observations' and analysis 
do not deal with scales larger than $2.5^\circ$. Fluctuation power below the
scale of correlations, $<\theta\sim 1^\circ-1.5^\circ$, for example due to galaxy 
cluster cores, does not affect the correlations significantly. This is illustrated
by \citet{dol97} and \citet{men02}, who have shown that including the fluctuation 
power on non-linear scales changes the amplitude of lensing-induced correlations 
by $\sim 10\%$, for scales $\simgt 10^\prime$.

We start with the assumption that without lensing GRBs and foreground APM galaxies 
are randomly distributed with respect to one another, on the sky. 
We now construct a {\it synthetic lensed} GRB subset. Suppose a large number
of GRBs with the BATSE-observed distribution of fluence and error properties go off 
in the direction of a patch with some $[n_{gal,D}/\langle n_{gal,R}\rangle]$ value,
picked at random from the solid line distribution in Fig.~\ref{subsample_hist}.
Using the various relations in Section~\ref{underden} and the fit in 
Fig.~\ref{overden}, we can predict the overdensity of GRBs in that direction:
\begin{equation}
q=
\Big(1-{\kappa_0\over b}
~\Big[{{n_{gal,D}}\over{\langle n_{gal,R}\rangle}}-1\Big]\Big)^{-2\beta}
\approx 
\Big(1+2\beta{\kappa_0\over b}
\Big[{{n_{gal,D}}\over{\langle n_{gal,R}\rangle}}-1\Big]\Big).
\label{q_full}
\end{equation}
The APM galaxies trace matter extending, at most, to $z=0.4$.  The corresponding 
optical depth in a flat $\Omega_m=0.3$ Universe is 0.025 for $z_s=1$, and 0.031 for 
$z_s=3$; we take $\kappa_0=0.028$, as an average. We set $b=1$, as estimated on 
large scales \citep{pea02}. We use $\beta=-0.62$ and $-0.45$, for the
46 and 74 synthetic GRBs subsets, as determined in Section~\ref{underden}). 
If the number of GRBs going off is 1, then $q$ is the probability that it will be
observed. Using the latter definition, a GRB is accepted into the synthetic
lensed subset with probability $q$; if $q>1$, the GRB is accepted into the 
subset, and an additional one is accepted with probability $q-1$. 
We repeat the process until we build up two separate subsets of 46 and 74 GRBs 
each; we repeat the procedure a 1000 times.

The average $[n_{gal,D}/\langle n_{gal,R}\rangle]$ in the 46 and 74 synthetic 
subsets is close to 0.996, barely below what one expects for a randomly selected 
set of points on the sky, and far from 0.883 and 0.915 which are found for the 
real subsets. In fact, the fraction of synthetic subsets whose average 
$[n_{gal,D}/\langle n_{gal,R}\rangle]$ is less than 0.883 and 0.915, is 0.3\% and 
0.75\%  respectively. Hence, this model fails to reproduce observations. 

How much do we have to change the parameters in eq.~\ref{q_full} to reach 
agreement with observations?
If $\kappa_0$ is increased by a factor of 10 (20), then the respective 
percentages for the 46 and 74 subsets become about 2\% (8\%), and 3\% (10\%),
so, roughly speaking, the observed subsets can occur with $~\simlt\, 10\%$ 
probability.
We conclude that observations could be deemed to agree with expectations if 
$\kappa_0\sim 0.4$. How realistic is this value? The optical depth of the entire 
column of matter between the observer and a source at $z_s=1.0$ (3.0) is 
$\kappa_0=0.065$ ($0.34$). It is very hard to imagine how APM galaxies, 
with faint galaxy magnitudes of around 20-21 in typical optical bands can be 
faithful tracers of matter fluctuations at redshifts 1-3. An alternative is to
keep $\kappa_0$ at 0.028, but require $b\sim 0.1$. This is inconsistent with  
observations that estimate $b$ on large scales to be within $\sim 30\%$ of unity 
\citep{pea94,gas01}. We will return to interpretation of our results in 
Section~\ref{we_can}.

\section{GRB--ACO cluster correlations}\label{ACO}

If nearby mass distribution is weakly lensing GRBs, then all tracers of the nearby 
mass should be anti-correlated with GRB events. Here we carry out cross-correlations 
between GRBs and   galaxy clusters from the Abell-Colwin-Olowin Catalog 
\citep{abe58}.

\subsection{Results from literature}\label{ACO_lit}

The present work is not the first to consider GRB-ACO correlations.
Prior to 1997, before the first spectroscopic redshifts to GRBs were established, 
several workers have looked for correlations between GRBs and ACO clusters.
Their motivation was that if GRBs are formed in galaxies, then the degree of their 
association with known populations of galaxy clusters will place constraints on 
their redshift distribution. All these studies tacitly assumed that weak lensing
effect was negligibly small. Based on the current cosmological models, and our 
predictions in Section~\ref{can_we?}, their assumption was perfectly justified.

Some studies detected correlations between subsets of clusters and 
subsets of GRBs, others detected no correlations. 
For example, \citet{kol96} claimed an association of 136
~$e\leq 1.6^\circ$ GRBs from the 3B Catalog and 3616 ~$|b|\geq 30^\circ$ ACO 
clusters at separations $\Theta\leq 4^\circ$, at a significance level of 95\%.
\citet{mar97} reported correlation of 71 ~$e\leq 1.685^\circ$ ~3B GRBs 
with 185 nearby ACO 
clusters with $R\geq 1$ and $D\leq 4$ at $2.9-3.5\sigma$ level. They found even 
stronger correlations between 27 ~$e\,\simlt\, 0.35^\circ$ ~3B GRBs and all 
5250 ACO clusters, at $3.5-4\sigma$ level. Given these findings, they were 
surprised that 40 very well localized Inter Planetary Network (IPN) GRBs were 
not correlated with ACO clusters. \citet{hur99} also used IPN positions 
but for a much larger sample, 157 GRBs from 4B Catalog. The average reduction 
in error area is $\sim 50$ for IPN compared to BATSE positions. Despite the 
accurate positions \citet{hur99} did not find any correlations, for assumed 
ACO cluster radii of $0.2^\circ$ or $0.4^\circ$. 

When GRB redshifts became available a few years ago, GRB-cluster
cross-correlation work stopped. In view of our APM galaxy results in 
Section~\ref{APM}, it becomes interesting to revisit GRB--ACO correlations, 
especially since the Current BATSE Catalog contains about twice as many GRB events
as were used in earlier studies, and error estimates for some GRB events
have been recently revised \citep{pac99}.

\subsection{Our results}\label{ACO_ours}

Abell-Colwin-Olowin Catalog clusters are classified into distance
and richness classes according to the standard Abell criteria \citep{abe58}.
Since lensing should be sensitive to cluster distances we divide the clusters 
into three distance ranges: nearby, D=1-4, which have an average redshift of 0.06, 
medium distant, D=5, with an average $z$ of 0.11, and distant, D=6, with an 
average $z=0.18$. Distance class 7 is very incomplete; it has 8 clusters vs. 
about 2000 D=6 clusters, so we leave out D=7 clusters from the analysis. 
We further divide the clusters into poor ones, R=0-2, and rich ones, R$\geq$3. 
ACO catalog is incomplete close to the Galactic plane, so we mask out areas 
at $|b|<30^\circ$. This leaves us with 3608 clusters and 1021 GRB events 
which we use to compute cross-correlations on a range of angular scales.

Fig.~\ref{ACO_corr_one} shows GRB-ACO correlation functions
between GRBs with $e\leq 1^\circ$ \footnote{This selection corresponds to the
74 GRB subset of the APM analysis, however, the number of GRBs here is actually
183. The difference arises mostly because in the APM analysis we did not use GRBs 
that were too close or too far from the plate center, whereas here there are no
such constraints.} and six subsets of ACO
clusters. 
We used \citet{landy93} estimator, which has improved variance compared to the 
standard estimator. The two sets of solid lines in each panel are the 95\% and 
99\% confidence limits derived from cross-correlations using 300 random 
realizations of the GRB catalog (compensated for the same declination dependent 
sky coverage as the real BATSE catalog). The dotted lines are the same 
correlations, but binned into narrower angular bins. (We do not plot confidence 
limits for these). 

Out of sixty points plotted in six panels of Fig.~\ref{ACO_corr_one} only one
is significant at $\geq 99\%$: it states that GRBs with $e\leq 1^\circ$ are
anti-correlated with rich distant clusters on scales $\simlt 20^\circ$. Of the 
six cluster subsets, the rich distant ones are most likely to act as weak lenses 
for background GRB, so this result supports our earlier GRB-APM findings. Whereas 
in the APM study the angular scales we could probe were limited by the plate size, 
here we are not restricted in that regard. We find that the full scale of 
anti-correlations corresponds to about 150$h^{-1}$Mpc at the redshift of the 
clusters, which is the scale where cluster power spectrum is observed to peak:
\citet{ret98} find peak $k=2\pi/l\sim 0.05 h$~Mpc$^{-1}$ for ACO clusters, while
\citet{tad98} find peak $k\sim 0.03 h$~Mpc$^{-1}$ for APM clusters.

Results with $0.5^\circ\leq e \leq 1^\circ$ (corresponding to the 46 GRB subset
of the APM analysis) are similar to the ones in Fig.~\ref{ACO_corr_one}, however,
the significance of anti-correlations with rich distant clusters drops to
$\sim 97\%$. Correlations with GRBs unrestricted in positional error show no 
signal.

\section{Discussion}\label{we_can}

Here we explore ways of accommodating GRB-APM anti-correlations, and earlier
reports of QSO-galaxy correlations \citep{wil98,nor00,bar97}. 
There are three possible avenues towards a resolution of the 
high amplitude of (anti-)correlations; these concern the sources (GRBs or QSOs), 
the lensing process, and the lenses (the mass distribution), respectively. 
We discuss these
in turn. We assume that in the weak lensing regime the effects of these three 
sets of possibilities add up `linearly', and so they can be considered separately.

\subsection{Sources}\label{sources}

Suppose our assumptions about GRB number distribution in fluence and error 
(Fig.~\ref{ncounts}) are incorrect, i.e. our assumed conversion between $M$ 
and $q$---exponent $\beta$ in eq.~\ref{q_full}---is wrong.
This could arise, for example, due to shot noise given small number of GRBs. 
The extreme value that $\beta$ can take is $-1$, which is when there are no
more sources beyond the limit of the survey, and area dilution of lensing
reduces the sky density of sources by $M$. Even this extreme case does not come
close to reproducing GRB-APM anti-correlations.

A similar solution to the problem was considered in \citet{wil98}, 
who observed positive correlations between bright, optically selected QSOs
and APM galaxies. The QSO number counts in the relevant redshift range
gave $q=M^{1.75}$, whereas $q=M^{\sim 20}$ would be required to reproduce
observations. That explanation was ruled out as unlikely.

\subsection{Lensing process}\label{lensing_process}

Is it possible that
propagation of light through an inhomogeneous Universe is not well 
modeled by the standard lensing equation? In terms of eq.~\ref{q_full} that would
mean that the expression in round parentheses is incorrect. 
This was considered in \citet{wil00}, who explored
the effect of the second order term in the lensing equation. Including 
that term had interesting consequences: magnification was increased,
but only by 10\%, and {\it only} in the mass distribution scenarios which were 
far from Gaussian random fields on large spatial scales.  So this does not 
appear to be a promising avenue for the resolution of the problem.
Furthermore, propagation of light through a numerically simulated Universe 
of popular cosmological models was done using the full geodesic equation of motion 
\citep{van01,tom99}, and the results were found to be 
in good agreement with the predictions based on the standard lensing equation.

\subsection{Lenses---constant biasing}\label{lenses1}

The last set of possibilities is that the mass fluctuations are quite different 
from the observed projected galaxy density fluctuations. To reproduce the GRB-APM 
anti-correlations, $\kappa_0/b$ in eq.~\ref{q_full}, would have to be increased
by a factor of 10-20. Increasing $\kappa_0$ by that large a factor is ruled out 
(Section~\ref{can_we?}). On the other hand, a constant biasing of factor of 0.1-0.05 
is ruled out by dynamical measurements on cluster and supercluster scales.

\subsection{Lenses---density dependent biasing}\label{lenses2}

Here we propose another variant of the third set of possibilities; we relax the 
requirement that $b$ is constant as a function of $\delta\sigma/\sigma$.
This is a toy model only. We keep $\kappa_0=0.028$, which is appropriate for
the average optical depth of the APM galaxies. Within this slab of 
APM galaxies, the projected galaxy density is assumed to be a monotonic, but not
linear tracer of the projected mass density. 

\begin{figure}
\plotone{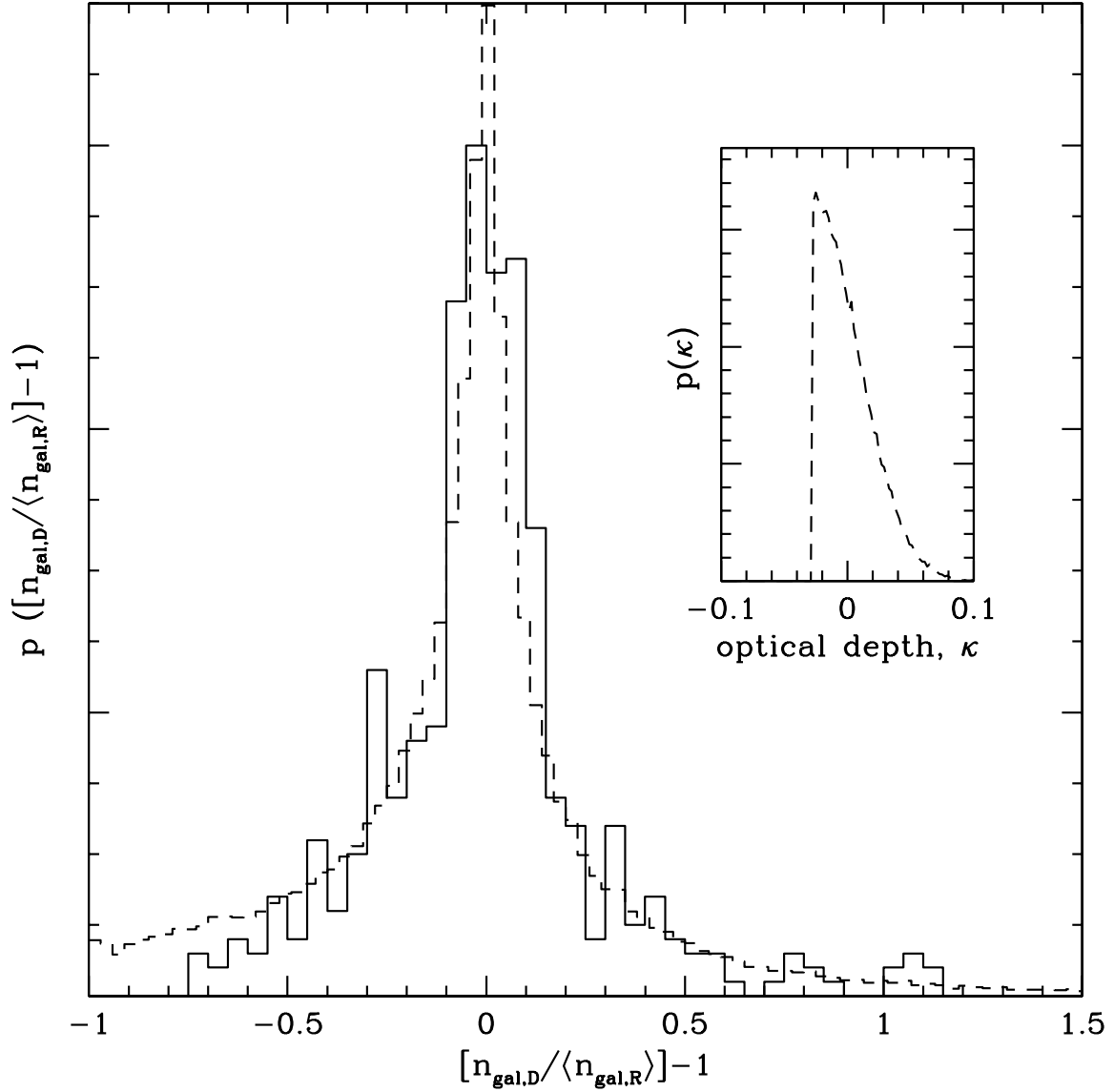}
\caption{Probability distribution of projected galaxy number density: solid
line is the APM counts-in-cells using circular cells of radius $\theta=1.5^\circ$,
while the dashed line is for the toy model proposed in Section~\ref{lenses2}.
The model has projected mass distribution (expressed in terms of lensing optical
depth $\kappa$) as shown in the inset, and biasing function as shown in 
Fig.~\ref{smmodel_bulk}. The model reproduces the overall shape of the APM 
counts-in-counts.
\label{smmodel_histo}}
\end{figure}

The shape of the distribution of APM galaxy counts-in-cells on $\theta=1.5^\circ$ 
scales is fixed by observations. To maximize lensing effects we chose a skewed 
shape for $p(\delta\sigma/\sigma)$:
let the distribution of projected mass densities averaged over circles of 
$\theta=1.5^\circ$ have the shape of a half-Gaussian, with the sharp cut-off 
coinciding with $\delta\sigma/\sigma=-1$, and the tail extending to positive 
values of $\delta\sigma/\sigma$. In terms of the optical depth the cutoff
is at $-\kappa_0$ (see insert in Fig.~\ref{smmodel_histo}); the corresponding
lines of sight are `empty beams' and produce maximum possible demagnification of 
the sources. The width of the half-Gaussian is adjusted such that the average 
$\delta\sigma/\sigma$, and hence the average $\kappa$ are zero.

\begin{figure}
\plotone{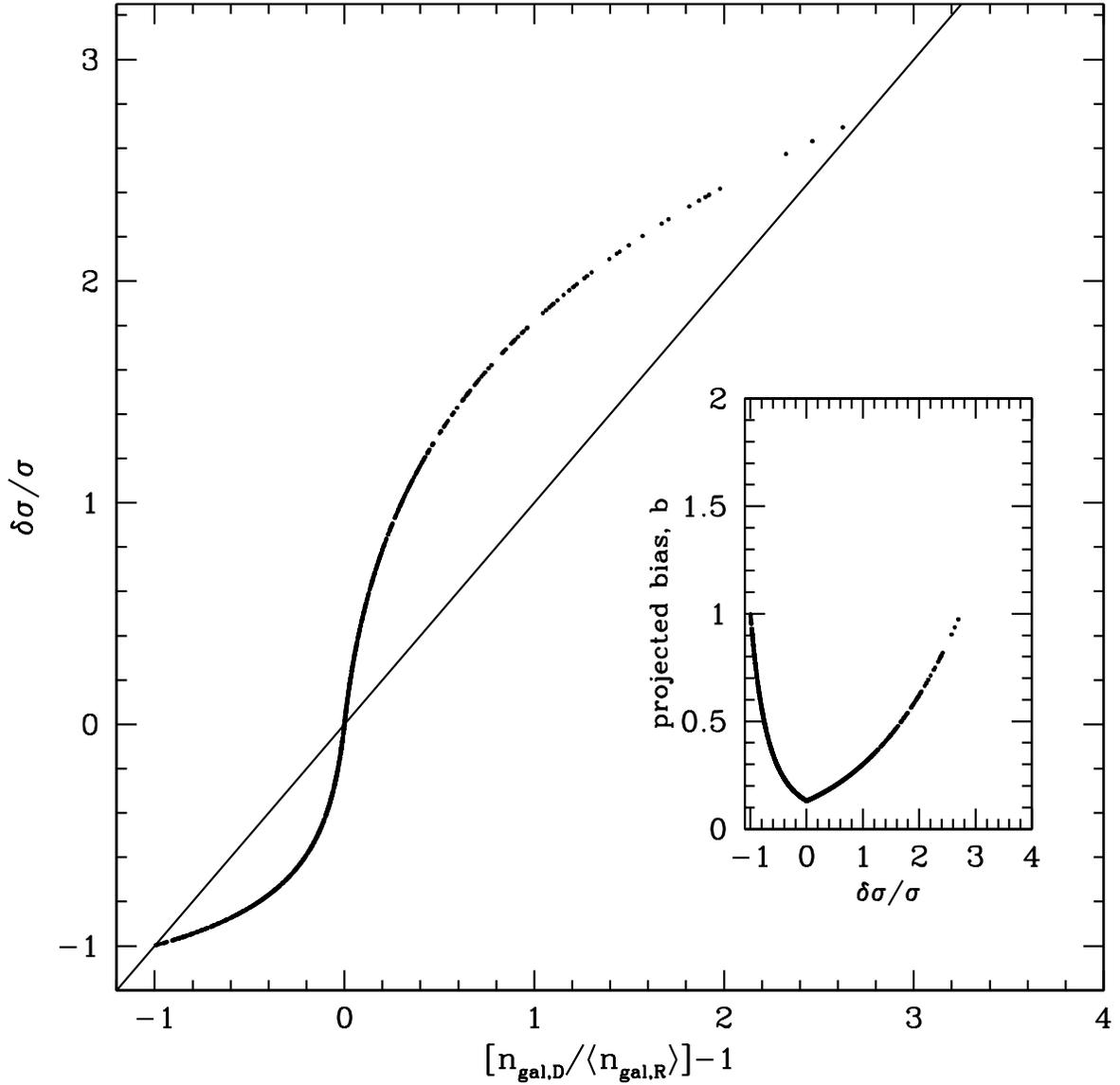}
\caption{
The projected biasing function, on scales $\sim 1.5^\circ$, for the toy model 
proposed in Section~\ref{lenses2}.
The main plot shows the relation between projected galaxy density excess and
projected mass excess, for the slab of matter between $z\approx 0.1$ and 
$z\approx 0.4$; the inset shows the biasing factor as a function of
projected mass excess. The straight line corresponds to $b=1$.
\label{smmodel_bulk}}
\end{figure}

\begin{figure}
\epsscale{0.95}
\plotone{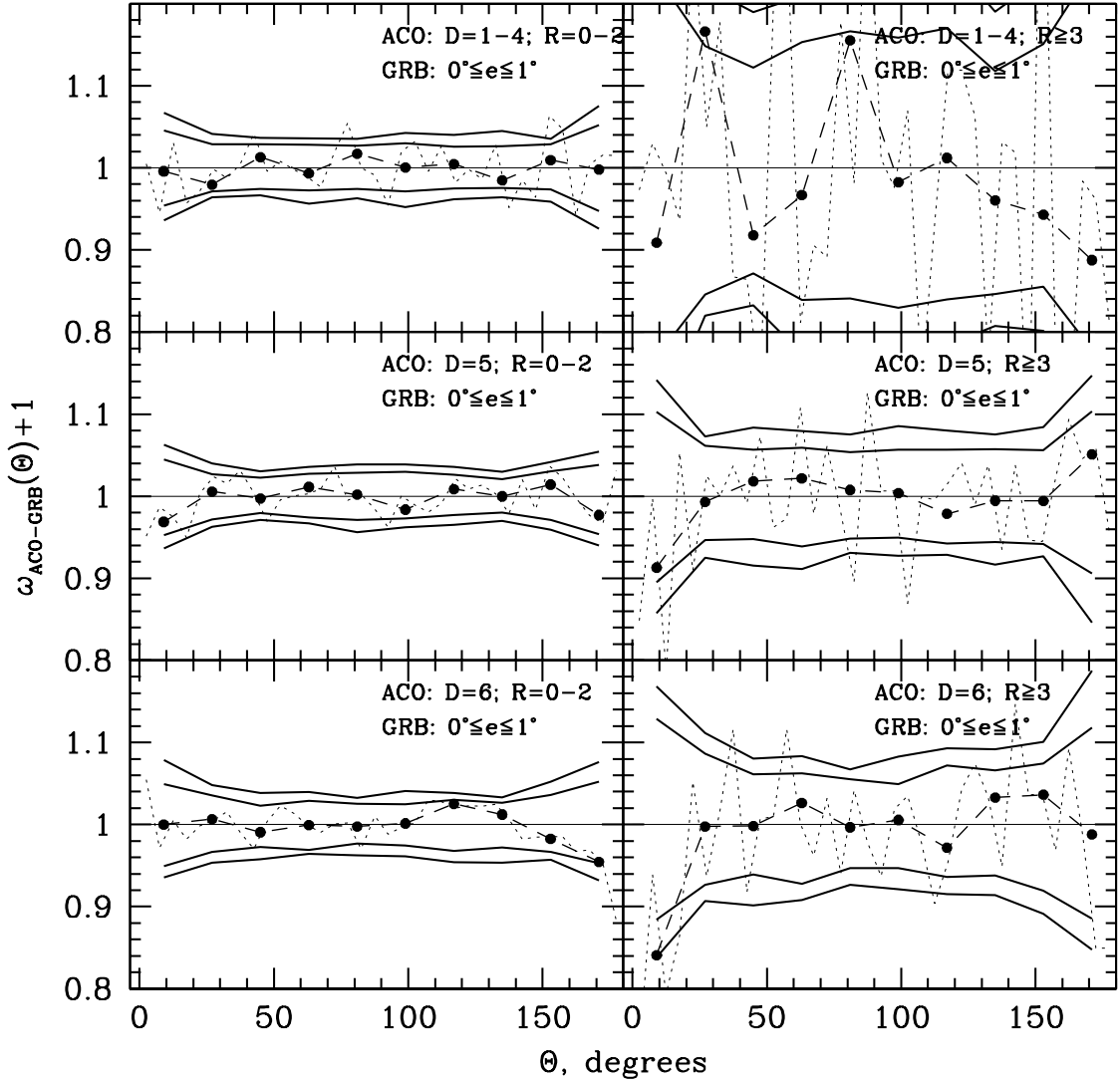}
\caption{
Angular cross-correlation between ACO clusters split up by distance and richness
class, and well localized BATSE GRBs, in $18^\circ$ bins. Because ACO catalog
is incomplete at low Galactic latitudes, clusters and GRBs are restricted to
$|b|\geq 30^\circ$. The two sets of solid lines in each panel are the
95\% and 99\% confidence limits derived from 300 random realizations of
GRB distributions (Section~\ref{ACO_ours}). The dotted lines are the same 
correlations, but binned into narrower angular bins. (No confidence limits are 
plotted for these). The only point significant at $\geq 99\%$ is the 
anti-correlation with rich distant clusters, i.e. the subset of clusters most 
likely to act as weak lenses for GRBs.
\label{ACO_corr_one}}
\epsscale{1.0}
\end{figure}

Since $p(\delta\sigma/\sigma)$ is highly asymmetric, while
the observed $p([n_{gal,D}/\langle n_{gal,R}\rangle]-1)$ is symmetric,
we must adjust the biasing function to map the former distribution onto the latter.
Fig.~\ref{smmodel_bulk} plots one possible version of such a biasing function, 
and the corresponding distribution of projected galaxy densities is shown as the 
dashed line in Fig.~\ref{smmodel_histo}. It is symmetric, and nearly zero-centered, 
i.e. similar to the observed APM distribution (the solid histogram in this figure is 
the same as the solid histogram in Fig.~\ref{smkstest2_histo}). 
In short, the biasing function takes the very skewed projected mass distribution 
(insert in Fig.~\ref{smmodel_histo}) and maps it onto a symmetric projected galaxy 
distribution (dashed histogram in Fig.~\ref{smmodel_histo}). 

Within this toy model,
the average galaxy density for the 46 and 74 synthetic GRB subsets is
about 0.97, and densities $\le 0.883$ ($\le 0.915$) for the 46 (74) subset 
occur in 9\% (13\%) of the cases. Thus the model can account for the observations.
Furthermore, this model also reproduces the observed correlations reported
in \citet{wil98}: 
if the slope of QSO number counts is $\alpha=1.1$ (as quoted in that study), 
our model gives 1.018 for the average normalized galaxy foreground density in 
circles $\theta=1.5^\circ$ around bright optically selected QSOs, 
a value consistent with Fig. 6 of that study ($\sim 1.01$).

Our model $p(\kappa)$ distribution and the resulting biasing function are very 
different from what is expected in currently accepted Universe models, with initial 
density perturbations specified by Harrison-Zel'dovich-type spectrum, and matter and 
cosmological constant (or the like) contributing comparably to the net zero 
curvature. In such models $p(\kappa)$ is Gaussian on large scales, and the 
biasing factor is 
close to unity for all values of density excess. For example, Fig. 13 of
\citet{jai00} show that the symmetric shape for $p(\kappa)$ is already attained at 
smoothing scale of $8^\prime$, and is symmetric for all larger scales.

Our model requires that there are lines of sight with projected radii 
$\sim 20 h^{-1}$~Mpc and extending from $z\sim 0.1$ to $0.4$, or 
$\sim 600h^{-1}$~Mpc in terms of proper length, that are nearly devoid of mass. 
At the same time, these lines of sight are not nearly as empty in terms of galaxies. 
Within standard theoretical framework, where primordial fluctuation
power spectrum and subsequent gravitational instability are solely responsible 
for structure on large scales, a dramatic redistribution of baryonic and dark 
matter, such as implied by our model, is not possible. 
Note however, that even though our model is astrophysically implausible, 
it does not violate the physical constraint that densities must remain
positive everywhere: the sharp cut-off in $p(\kappa)$
corresponds to $-\kappa_0$, the minimum possible lensing optical depth presented 
by mass between $z=0.1$ and $0.4$ to sources at $z_s\approx 2.5$, in a
flat $\Omega=0.3$ universe model. 

Since baryonic and dark matter are unlikely to be distributed differently on 
large scales, we propose an alternative, speculative scenario.  
Suppose $\sim 70\%$ of the present day closure density is contributed by some
dark energy that, because of its equation of state became dynamically 
important only recently, say $z\,\simlt\, 1$. 
Until $z\sim 1$ structure formation proceeded 
according to the standard picture, with galaxies tracing the total mass, and with
galaxy and mass distributions looking Gaussian on large scales.
Suppose also that the dark energy can cluster on sub-horizon scales,
or, more specifically, on scales as small as $\sim 100-200 h^{-1}$ Mpc.
Then at late epochs the dark energy will start contributing to the potential 
fluctuations that are already defined by dark and baryonic matter, on those scales.
If these fluctuations develop quickly, then matter (light and dark), 
hampered by inertia, would not be able to respond quickly. Thus, power spectra 
obtained from galaxy surveys would not betray anything unusual.
On the other hand, light rays from distant sources will traverse the fluctuations at 
$c$, thereby `capturing' the total amplitude of the fluctuations. So, lensing is 
able to probe the full extent of gravitational potential wells, and if these are
deep, then significant source-lens (anti-)correlations would result.

In this picture, $p(\kappa)$ distribution introduced earlier would refer not to
the dark matter, but mainly to the projected clumping of dark energy. 
Assuming 3D clustering scale of $\sim 100-200 h^{-1}$ Mpc, lines of sight 
$\sim 600h^{-1}$~Mpc long that are devoid of dark energy are possible. 
Given the speculative 
nature of the proposed scenario, we will not develop it any further. Instead,
we ask if the qualitative features of our model are compatible with 
the existing observations. By qualitative features we mean the relation between
fractional galaxy excess and fractional dark energy excess 
(Fig.~\ref{smmodel_bulk}), with biasing now redefined in terms of these two 
quantities.
Specifically, do observations rule out strong anti-biasing 
for regions with small ~$|\delta\sigma/\sigma|$, biasing for regions with 
$\delta\sigma/\sigma\simgt\,$few, and moderate anti-biasing or $b\sim 1$ for all 
other regions. In 3D, this general behavior of the biasing function will be same as 
in projection. 

\vvskip
{\it Cosmic Microwave Background.}
According to the proposed scheme the dynamical evolution at very high redshifts
is same as in the standard cosmological scenarios, so the primary CMB fluctuations 
would remain unaltered. The dark energy fluctuations grow at $z\,\simlt\, 1$, 
and are 
therefore expected to contribute to CMB as late integrated Sachs-Wolfe effect, at 
$\ell\approx 180^\circ/\theta\sim 100-200$. This location coincides with that of
the first acoustic peak, and so the ISW signature will be either masked by the peak, 
or enhance the peak's amplitude compared to that of the secondary peak.
This is not ruled out by CMB observations.

\vvskip
{\it Galaxy redshift surveys.} 
Because of the late emergence of dark energy fluctuations, and galaxies' 
sluggishness in responding to these, the observed galaxy distribution at 
redshifts $\simlt\, 0.5-1$ should be similar to what one would expect in 
standard cosmological models.

\vvskip
{\it Cosmic velocity flows.}
Regions with $|\delta\sigma/\sigma|\,\simlt\,1$  would show significant 
anti-biasing according to our model. Such regions within a few hundred Megaparsecs 
around us are studied by cosmic velocity flows, and in the linear regime yield 
values for $\Omega_m^{0.6}/b$ \citep{dek94}. There are three basic methods of 
reconstructing 3D mass density fields from data: using density-density comparison, 
velocity-velocity comparison, or redshift-space distortions. These methods often
produce discrepant results for $\Omega_m^{0.6}/b$, with density-density giving 
consistently higher values than velocity-velocity comparison \citep{ber01}. 
For example, IRAS 1.2Jy + Mark III data set yields 
$\Omega_m^{0.6}/b=0.89\pm 0.12$ using density-density technique
\citep{sig98}, while $\Omega_m^{0.6}/b=0.50\pm 0.04$ is obtained 
using velocity-velocity comparison \citep{wlk98}. 
Other data sets produce similar results.
\citet{ber01} argue that no reasonable biasing scheme would generate 
such different $\Omega_m^{0.6}/b$ emerging from different methods, and 
ascribe the discrepancy to errors. This may well be the case,
however, the discrepancy could also arise if biasing is a strong function
of total density excess, as proposed by our toy model.
In that case, density-density comparison would yield most reliable results,
because the method makes no assumptions about the underlying mass density field,
but derives it from observed velocities.
Density-density comparisons currently yield $\Omega_m^{0.6}/b\sim 1$ 
\citep{sig98}; for $\Omega_m\sim 0.3$ this would imply $b\sim 0.5$, not too 
different from the average biasing factor of our proposed model, which is 0.62.
Thus our model is not necessarily in conflict with observed cosmic velocity flows. 

\vvskip
{\it Cosmic weak shear lensing.}
Another important observational test of our model is provided by the recent 
measurements of weak shear \citep{bac00,hoe02,kai00,mao01,van00,wit00}, 
which are consistent with $\Omega_m=0.3$ flat cosmological models, 
$b\sim 1$\footnote{To be more precise, \citet{mel01} derive an empirical fit:
$\Omega_m^{0.47}\sigma_8\approx 0.59_{-0.03}^{+0.03}$.}, and Gaussian mass 
distribution on large scales. Even though the total mass and dark energy 
distribution in our model is far from Gaussian, our model and weak shear 
observations may not be discrepant; we list three possibilities:
{\it (i)} The scales of weak shear and GRB-APM anti-correlations 
are different by about a factor of 3: the former apply to $\Theta<30^\prime$,
while the latter are on scales $\simlt 1.5^\circ$, thus the two techniques 
probe different physical scales. 
{\it (ii)} Observational factors may play a role. The redshifts of individual 
galaxies in the weak shear studies are not known, and so the lenses cannot be 
cleanly separated from the sources. Because of that, weak shear method may 
suffer from a `signal dilution' effect: 
a foreground overdense patch will produce coherent shearing in background
galaxies, but will also dilute the shear signal because of the increased
number of foreground unsheared galaxies.
{\it (iii)} In models with Gaussian distributed matter rms values of convergence 
and shear are the same. Suppose the distribution of dark energy trapped in potential 
wells of large overdense regions is top-hat-like. In that case the rms of 
convergence 
can be a few times larger than rms of shear, when averaged over large portions of 
the Universe. Thus the current weak shear observations may be underestimating the 
total amplitude of fluctuations.

We conclude that the qualitative features of our proposed model are not strongly 
ruled out by observations.

\section{Summary and Discussion}\label{sum+disc}

As discussed in the Introduction, bright optically-selected and radio-selected 
QSOs are known to be correlated with foreground galaxies on
large angular scales. The statistical significance of any one of these
studies is generally $2-3\,\sigma$, and can be dismissed as a chance occurrence, 
but collectively these studies imply that the correlations are real,
in which case the most likely explanation is weak magnification lensing.
The purpose of this work was to see if cosmologically distributed sources
other than QSOs are affected by weak lensing. 

We started by computing the expected magnification bias of GRB events due to
weak lensing, and found that most GRB subsets limited by statistical positional
error should show negative magnification bias, i.e. should be anti-correlated
with the foreground mass. Using APM galaxies to trace the mass we then
looked for any correlations on the sky between APM galaxies and GRBs. Because the 
projected number density of GRBs is low, about 1 per APM plate, we did not use 
cross-correlation analysis of counting pairs, but instead counted galaxies in 
circles around individual GRBs. We found that GRBs with small positional errors
are preferentially located in the directions where APM galaxies show $\sim 10\%$ 
deficits on degree angular scales. In particular, a subset of 46 GRBs with 
$0.5^\circ \le\,e\le\,1^\circ$ and a subset of 74 GRBs with 
$0^\circ \le\,e\le\,1^\circ$ have 12\% and 8\% less galaxies in circles of
$1.5^\circ$ radii around them (or $\sim 20 h^{-1}$ Mpc at the redshift of the
typical lenses), than expected on average. These are significant at 99.7\% and 
97.6\% confidence levels. Whereas these significance levels are less than
overwhelming, what makes the finding especially interesting is that the GRBs are 
{\it anti-correlated} with the APM galaxies, as expected from the magnification 
bias (and opposite to what is expected if physical associations are significant). 
This is the first reported case of weak lensing induced anti-correlations on degree 
angular scales.

To test the observed anti-correlation of GRBs with foreground mass, we carried 
out a cross-correlation analysis between GRB subsets classed by positional error
and Abell-Corwin-Olowin galaxy clusters of three distance ranges, D=1-4, D=5, and D=6,
and two richness subsets, R=0-2, and R$\ge 3$. 
Only one combination of GRBs and clusters showed significant results: 
$0^\circ \le\,e\le\,1^\circ$ GRBs and rich distant ACO clusters are anti-correlated
on $\simlt 20^\circ$ scales, consistent with predictions in Section~\ref{underden},
and observed anti-correlations with APM galaxies.

As was the case with earlier reports of QSO-galaxy correlations, present
anti-correlations with GRBs cannot be accounted for by weak lensing using a 
standard model of mass distribution and moderate biasing on large scales; 
the discrepancy is rather severe, one needs to increase the lensing optical 
depth by 10, or reduce the biasing factor to $b\sim 0.1$, which is not a viable 
option. Since both the GRB-APM and GRB-ACO anti-correlation signals are of 2-3 
$\sigma$ significance, they could well be statistical flukes. However, if 
they are not, and if the signal is due to weak lensing, then there are three 
types of resolutions: those that have to do with sources, lensing process, or 
lenses. We argue that the first two are inadequate, leaving us with the third. 

We propose a speculative scenario which assumes that $\sim 70\%$ of closure density 
is contributed by dark energy that can clump on significantly sub-horizon scales, 
$\sim\,100-200 h^{-1}$ Mpc, and that this clumping developed recently, 
$z\,\simlt\, 1$. Then at present epochs there should be significant fluctuations in 
the dark energy component. However, the amplitude of baryonic and 
dark matter fluctuations would remain relatively unchanged, because it would take 
time for matter to be accelerated and displaced by significant amounts on large 
spatial scales. In this scenario, the galaxy distribution would trace the 
underlying potential wells but would severely underestimate the amplitude of 
fluctuations on $\simgt 100 h^{-1}$ Mpc scales. Dynamical measurements would do 
better, since velocities `respond' quicker to acceleration than displacements. 
Lensing, which relies on light traversing the fluctuations at $c$, is the best 
tool to probe the full extent of these fluctuations. In this scheme, weak lensing 
induced GRB-APM anti-correlations would be strong because the total amplitude of 
gravitational potential fluctuations are substantially larger than 
galaxies would have us believe. 

In principle, weak shear lensing should be able to detect the true extent of 
these fluctuations as well. However, weak shear observations support the standard 
cosmological model with cold dark matter and unclustered dark energy component, 
and seem to be in conflict with our proposed scenario. Current observations of 
microwave background anisotropy, galaxy redshift surveys and cosmic velocity flows 
are not strongly incompatible with our hypothetical scenario.

%%\clearpage

\clearpage

\begin{deluxetable}{rccccc}
\tablecolumns{6}
\tablewidth{0pt}
\tablecaption{GRB subsets that show significant
anti-correlations with APM galaxies.}
\tablehead{
\colhead{$\theta$}& $d_{cen}$ & $e$   & \# GRBs in & total \# & Signif. \\
\colhead{}        & range     & range & subset  & of GRBs  & \% 
}
\startdata
         0.5 & 1.5--2.5  &  0.5--1.0 &  46 & 448 &  98.01 \\
         1.0 & 0.0--2.5  &  0.5--1.0 &  88 & 732 &  97.53 \\
         1.0 & 0.5--2.5  &  0.5--1.0 &  81 & 689 &  97.15 \\
         1.0 & 1.5--2.5  &  0.5--1.0 &  46 & 448 &  99.63 \\
         1.0 & 1.5--2.5  &  9.5-10.0 &   6 & 448 &  97.44 \\
 $\star$ 1.5 & 1.5--2.5  &  0.5--1.0 &  46 & 448 &  99.69 \\
 $\star$ 1.5 & 1.5--2.5  &  0.0--1.0 &  74 & 448 &  97.62 \\
\enddata
\end{deluxetable}

\end{document}